\documentclass[a4paper,11pt]{article}%,centerh1

\usepackage[sans]{dsfont}
\usepackage[mathscr]{euscript}

\usepackage[centertags]{amsmath}
\usepackage[nosort]{cite}
\usepackage[psamsfonts]{amsfonts}
\usepackage{amsthm,enumerate,amssymb}

\newcommand{\rmd}{\mathrm{d}}
\newcommand{\rmi}{\mathrm{i}}
\newcommand{\rme}{\mathrm{e}}
\newcommand{\openone}{\mathds1}
\newcommand{\norm}[1]{\left\Vert#1\right\Vert}
\newcommand{\abs}[1]{\left\vert#1\right\vert}
\newcommand{\T}{\mathtt{T}}
\newcommand{\Rbb}{\mathbb{R}}
\newcommand{\Cbb}{\mathbb{C}}
\newcommand{\Pbb}{\mathbb{P}}
\newcommand{\Qbb}{\mathbb{Q}}
\newcommand{\Ebb}{\operatorname{\mathbb{E}}}

\newcommand{\RE}{\operatorname{Re}}
\newcommand{\IM}{\operatorname{Im}}
\newcommand{\Tr}{\operatorname{Tr}}

 \newcommand{\Dcal}{\mathcal{D}}
 \newcommand{\Ecal}{\mathcal{E}}

\newcommand{\Lcal}{\mathcal{L}}

\newcommand{\Scal}{\mathcal{S}}

\newcommand{\Fscr}{\mathscr{F}}

\newcommand{\Hscr}{\mathscr{H}}

\newcommand{\Kscr}{\mathscr{K}}
\newcommand{\Lscr}{\mathscr{L}}

\newcommand{\Sscr}{\mathscr{S}}

\newcommand{\Uscr}{\mathscr{U}}

\newcommand{\ms}{\mathfrak{Z}}
\newcommand{\K}{\mathtt{K}}

\begin{document}

\title{Entanglement protection and generation under continuous monitoring}

\author{A. BARCHIELLI and M. GREGORATTI
\\{\footnotesize \textsl{Politecnico di Milano, Department of Mathematics ``F.Brioschi''}} \\
{\footnotesize \textsl{Piazza Leonardo da Vinci 32, I-20133 Milano, Italy}} \\
{\footnotesize \textsl{Also: Istituto Nazionale di Fisica Nucleare, Sezione di Milano}}}

\date{}

\maketitle

\begin{abstract}
Entanglement between two quantum systems is a resource in quantum information, but dissipation usually destroys it. In this article we consider two qubits without direct interaction and we show that, even in cases where the open system dynamics destroys any initial entanglement, the mere monitoring of the environment can preserve or create the entanglement, by filtering the state of the qubits. While the systems we study are very simple, we can show examples with entanglement protection or entanglement birth, death, rebirth due to monitoring.

\end{abstract}

\section{Introduction}
Entanglement is an intrinsically quantum type of correlation among quantum systems which is of fundamental importance in quantum information\cite{NC00}. The behaviour of entanglement under dissipative dynamics has been studied extensively\cite{Ors11,YuE09}, either to find means to protect entanglement against decoherence, either to understand how to use a dissipative dynamics to create entanglement. Usually dissipation tends to destroy entanglement, at least when the two quantum systems do not interact directly. Sometimes this disentaglement can be completed even in a finite time\cite{Dio03,YuE07,Ors11} and this phenomenon has been called \emph{entanglement sudden death} (ESD). However, dissipation can create entanglement too; this happens when the two parties interact with a common bath\cite{BFM09,BFM10,BLP10,YuE09,Ors11}, even if they do not interact directly, and we can have entanglement birth, death, rebirth. Entanglement can be preserved or generated also by controlling the composite system by means of measurement based feedback\cite{CarvH07,CRH08,MMCTF10}.

Preservation of entanglement can be obtained also by pure monitoring of the system\cite{VogS10,VGCBB10,MCVF11}, that is by an indirect measurement on the system which acquires information thanks to the observation of its environment, but which does not perturb the system. Quantum trajectory theory allows for describing a continuous monitoring\cite{Bar06,BarGreg09} and in such a theory we have to distinguish between the \emph{a posteriori state}, the conditional state given the observed output, and the \emph{a priori state}, the mean state, satisfying a master equation. It is possible that the a posteriori states are entangled, while the a priori state is not. By using the \emph{concurrence}\cite{Woo98} as a measure of entanglement it has been shown that the pure monitoring can slow down the decay of the entanglement\cite{VogS10}.

The aim of our paper is indeed to study the effect of monitoring on the a posteriori entanglement when the a priori dynamics washes out any initial entanglement. More precisely, we consider the case of the open dynamics of two qubits in the Markovian regime and we model their global evolution by a Hudson-Parthasarathy equation. This approach allows to clearly characterize the Markovian evolutions representing two qubits which do interact or do not interact, directly or through a common bath. Section \ref{sec:QSDE} is devoted to the HP evolutions and to such a characterization; we recall also how to introduce measurements continuous in time and how to get the corresponding \emph{stochastic Schr\"odinger equation} (SSE) and \emph{stochastic master equation}, which are the starting points to study the dynamical behaviour of the monitored system and of its entanglement.
In Section \ref{sec:noint} we consider the case of no direct or indirect interaction between two qubits. When only local detection operators are involved, we show that, by pure monitoring, the decay of entanglement can be slowed down and, in special cases, even stopped independently of the qubit initial state (entanglement protection). In cases with non local detection operators, we show that, now depending on the qubit initial state, entanglement can even be created by pure monitoring (entanglement generation). In Section \ref{sec:jump} we study a case of indirect interaction between the two qubits through a common bath. We show  that, even if the a priori dynamics completely destroys any entanglement, a proper monitoring scheme can maximally entangle any initial qubit state.

\subsection{Two qubits}

We consider two qubits; for each qubit we denote by $|1\rangle$ the \emph{up} state and by $|0\rangle$ the \emph{down} state. By $\sigma_x$, $\sigma_y$, $\sigma_z$ we denote the Pauli matrices. In $\Hscr=\Cbb^2\otimes \Cbb^2$ the canonical basis (or  \emph{computational basis})\cite{NC00} is
\begin{equation}\label{cbas}
|u_1\rangle =|11\rangle, \quad |u_2\rangle =|10\rangle, \quad |u_3\rangle
=|01\rangle, \quad |u_4\rangle =|00\rangle,
\end{equation}
and the \emph{Bell basis}\cite{Petz} is %\cite{NC00}
%\begin{equation}\label{Bellb}\begin{split}
%&|\beta_0 \rangle = \displaystyle \frac{1}{\sqrt{2}}\left(|00\rangle +
%|11\rangle\right),
%    	\quad
%|\beta_1\rangle = \sigma_x\otimes \openone|\beta_0\rangle=\displaystyle \frac{1}{\sqrt{2}}\left(|10\rangle +
%|01\rangle\right),
%			\\
%&|\beta_3 \rangle = \displaystyle \frac{1}{\sqrt{2}}\left(|00\rangle -
%|11\rangle\right),
%\quad
%|\beta_4 \rangle =\frac{1}{\sqrt{2}}\left(|01\rangle-|10\rangle \right).
%\end{split}
%\end{equation}
\begin{equation}\label{Bellb}
|\beta_0 \rangle =  \frac{1}{\sqrt{2}}\left(|00\rangle +
|11\rangle\right),
    	\qquad
|\beta_i\rangle = \sigma_i\otimes \openone|\beta_0\rangle, \quad i=1,2,3.
\end{equation}

The set of \emph{statistical operators} is $\Sscr(\Hscr)$ and the one of linear operators is $\Lscr(\Hscr)$. A \emph{local operator} is a linear operator which acts non trivially only on one of the factors of $\Cbb^2\otimes \Cbb^2$, i.e.\ it has the form $A\otimes \openone$ or $\openone\otimes  A$ with $A\in \Lscr(\Cbb^2)$. The two qubits are independent if their state is a product state $\rho=\rho_1\otimes \rho_2$. The \emph{separable states}\cite{HHH96} are the statistical operators which admit a convex decomposition into product states, so that the correlation between the two qubits has a classical explanation; the other statistical operators are said to be \emph{entangled}. The \emph{maximally entangled states} are the pure states which, by partial trace on one of the two factors, reduce to maximally chaotic states, that is $\openone/2$. The projection on one of the Bell vectors \eqref{Bellb} is a maximally entangled state.

\subsection{Concurrence}\label{sec:conc}

A very useful measure of entanglement is the \emph{concurrence}, introduced by Wootters\cite{Woo98}. Let us consider a generic vector $\varphi\in\Hscr$ and expand it on the canonical basis \eqref{cbas}
\begin{equation}
\varphi = \varphi_{11}|11\rangle + \varphi_{10}|10\rangle + \varphi_{01}|01\rangle +\varphi_{00}|00\rangle.
\end{equation}
Let $\T$ be the complex conjugation of the coefficients in the canonical basis:
\begin{equation}
	\label{Tconj}
\T \varphi = \overline{\varphi_{11}} \,|11\rangle + \overline{\varphi_{10}} \,|10\rangle +
\overline{\varphi_{01}} \,|01\rangle +\overline{\varphi_{00}} \,|00\rangle.
\end{equation}
Let us define
\begin{equation}\label{Cphi}
\chi_{\varphi} : = \langle \T \varphi | \sigma _y \otimes \sigma _y \varphi \rangle
=2 \left(\varphi_{10}\varphi_{01} - \varphi_{11}\varphi_{00}\right), \qquad C_{\varphi} : = \left |  \chi_{\varphi} \right |.
\end{equation}
When $\norm{\varphi}=1$, $C_{\varphi}$ is the \emph{concurrence} of the pure state
$\varphi$. In general, if $\varphi$ is not normalized and $\psi =
\frac{\varphi}{\norm{\varphi}}$, then
\begin{equation}\label{Cpsi}
C_{\psi} = \frac{C_{\varphi}}{\norm{\varphi}^2}.
\end{equation}
%By using the Bell basis, from $\varphi= \sum_{i=0}^3 \phi_i|\beta_i\rangle$, we get
%\begin{equation}\label{CBell}
%\chi_\varphi= {\phi_1}^2+{\phi_2}^2+{\phi_3}^2-{\phi_0}^2.
%\end{equation}
Note that $C_{\beta_j}=1$ and $C_{u_j}=0$.

If $\rho$ is a generic statistical operator, the concurrence is defined by
\begin{equation}\label{Cmixed}
C_{\rho}: = \inf \sum_{i}p_i C_{\psi _{i}},
\end{equation}
where the infimum is taken over all decompositions of $\rho$ in pure states,  $\rho
= \sum_{i}p_i | \psi _i\rangle \langle \psi _i|$, see for instance [\citen{Ors11}] p.\
231. We have $0\leq C_\rho\leq 1$, $\forall \rho \in \Sscr(\Hscr)$, with $C_\rho=0$ if and only if $\rho $ is separable and $C_\rho=1$ if and only if $\rho $ is maximally entangled.

A subclass of states, for which it is easy to compute the concurrence, is the one of the ``X'' states
\cite{YuE07,Ors11}: in the canonical basis, an X state has non vanishing matrix
elements only in the two main diagonals. The projection on a Bell vector is an X
state.
For any X state $\rho$, by setting $\rho_{ij}=\langle u_i|\rho u_j\rangle$, we have $\rho_{jj}\geq 0$,
$\rho_{ij}= \overline{\rho_{ji}}$, $\sum_{j=1}^4 \rho_{jj}=1$, $\rho_{11}\rho_{44}\geq \abs{\rho_{14}}^2$,
$\rho_{22}\rho_{33}\geq \abs{\rho_{23}}^2$; moreover, the concurrence is given by
\cite{YuE07}
\begin{subequations}\label{CX}
\begin{gather}
C_\rho=2\max \left\{ 0,\, C_1,\, C_2 \right\},
\\
C_1=\abs{\rho_{23}}-\sqrt{\rho_{11}\rho_{44}}, \qquad C_2=\abs{\rho_{14}}-\sqrt{\rho_{22}\rho_{33}}.
\end{gather}
\end{subequations}

Finally, let $A$ and $B$ be linear operators on $\Cbb^2$. In studying the dynamics of the concurrence, the following formulae will be very useful:
\begin{subequations}\label{ABX}
\begin{gather}
%\chi_{(A\otimes \openone)\varphi}=\chi_{(\openone\otimes A )\varphi}=
%\left({\det}_{\Cbb^2} A\right) \chi_\varphi, \qquad
\chi_{(A\otimes B)\varphi}=
\left({\det}_{\Cbb^2} A\right) \left({\det}_{\Cbb^2} B\right) \chi_\varphi,
\\
\langle \T \varphi|(\sigma_y A)\otimes \sigma_y \varphi \rangle =
\langle \T A\otimes \openone\varphi|\sigma_y \otimes \sigma_y \varphi \rangle =
\frac 1 2 \left( {\Tr}_{\Cbb^2} A\right) \chi_\varphi.
\end{gather}
\end{subequations}

\section{Global evolution and continuous measurements}\label{sec:QSDE}
The way to understand whether the two qubits interact or do not interact, directly or indirectly, is to look at the unitary dynamics of the two qubits plus their environment. In the Markov regime this can be done by starting from a quantum stochastic differential equation \`a la Hudson and Parthasarathy (HP equation)\cite{Parthas92} and this is also a clear way to introduce continuous mesurements\cite{BarP96,Bar06}.

As before the system space is $\Hscr$, while we take as environment space the symmetric Fock space $\Kscr=\Gamma[L^2(\Rbb;\ms)]$; $\ms$ is a complex Hilbert space, which will be only finite dimensional in the present paper. Let $U_t=\rme^{-\rmi t H_T}$, $H_T=H_T^*$, denote the unitary (Hamiltonian) global evolution in $\Kscr\otimes\Hscr$. We suppose that the free environment evolution is $\Theta_t=\rme^{-\rmi tE_0}$, the second quantization of the left shift, with its free Hamiltonian $E_0$. Then the global evolution in interaction picture with respect to $\Theta_t$ is
$$V(t) = \Theta_t^*\,U_t = \rme^{\rmi E_{\scriptscriptstyle 0} t}\,\rme^{-\rmi tH_T}, \qquad \qquad t\geq0,$$
which, in the Markov regime, can be defined directly by a HP-equation.

\subsection{HP evolutions}
We fix a basis $\{|z\rangle\}_{z\in Z}$ in the Hilbert space $\ms$. Let $a_z(t)$ and $a_z^\dagger(t)$ be the fundamental Bose field operators in $\Gamma[L^2(\Rbb;\ms)]$ and $A_z(t)=\int_0^t a_z(s)\rmd s$, $A_z^\dagger(t)=\int_0^t a_z^\dagger(s)\rmd s$, $\Lambda_{zw}(t)=\int_0^t a_z^\dagger(s)a_w(s)\rmd s$ be the fundamental integrators of quantum stochastic calculus.

Let us consider the HP-equation\cite{Parthas92} for unitary operators on $\Kscr\otimes \Hscr$
\begin{multline}\label{VHPeq}
\rmd V(t)=\biggl[\sum_{z,w\in Z}\left(S_{zw}-\delta_{zw}\right)\rmd \Lambda_{zw}(t)
- \sum_{z,w\in Z}L_z^*S_{zw}\,\rmd A_{w}(t)\\ {} + \sum_{z\in Z}L_z\,\rmd A^\dagger_z(t)
- \rmi H \rmd t-\frac{1}{2}\sum_{z\in Z}L_z^*L_z\,\rmd t\biggr]V(t);
\end{multline}
the initial condition is $V(0)=\openone$. By taking
\begin{enumerate}
\item $H,\, L_z,\, S_{zw} \in \Lscr(\Hscr)$ (bounded operators), $\forall z,w\in Z$,
\item $H=H^*$,
%\item $L\in\Lscr(\Hscr;\ms\otimes\Hscr)$, where $L h = \sum_z |z\rangle\otimes L_zh$,
\item $S\in\Uscr(\ms\otimes\Hscr)$ (unitary operators), where $ S = \sum_{zw}
    |z\rangle\langle w| \otimes S_{zw} $,
\end{enumerate}
the solution of \eqref{VHPeq} is indeed unique and unitary. Every operator is identified with its natural extension to $\Kscr\otimes\Hscr$.

By using the time ordered exponentials introduced by Holevo\cite{Hol96}, the solution $V(t)$ can be represented as
\begin{multline}\label{solHP}
V(t)=\overleftarrow{\exp}\Biggl\{-\rmi\int_0^t\biggl[\sum_{zw}K_{zw}a^\dagger_z(s)a_{w}(s)
- \sum_{zw}L^*_z\Bigl(\frac{K}{\openone-S^*}\Bigr)_{zw}a_{w}(s) \\
{}+ \sum_{zw}\Bigl(\frac{K}{S-\openone}\Bigr)_{zw}L_za^\dagger_{w}(s)
+ H + \sum_{zw}L^*_z\Bigl(\frac{K-\sin K}{4\big(\sin(K/2)\big)^2}\Bigr)_{zw}L_{w}\biggr]\rmd s\Biggr\},
\end{multline}
where
$S=\rme^{-\rmi K}$, with a selfadjoint operator $K$ on $\ms\otimes\Hscr$.

Moreover,
%if $\Theta_t$ is the second quantization of the left shift in $L^2(\Rbb;\ms)$
we have that $U_t$, defined by $U_t:=\Theta_tV(t)$ for $t\geq 0$, and by $U_t:=U_{-t}^{\;*}$ for $t\leq 0$, is a unitary strongly continuous group. So, we can interprete $U_t$ as the evolution operator of a closed system, $\Theta_t$ as the free evolution of the fields and $V(t)$ as the total evolution in the interaction picture with respect to $\Theta_t$.

The interaction between $\Hscr$ and $\Kscr$ is regulated by the system operators $H$, $L_z$ and $S_{zw}$; the corresponding global Hamiltonian $H_T$ is a very singular unbounded operator which could even encode the whole interaction just in the shape of its domain\cite{Greg01}. Anyway, thanks to representation \eqref{solHP}, the global Hamiltonian $H_T$ has the heuristic expression
\begin{multline}\label{heuristichamiltonian}
H_T= E_0 + \sum_{zw}K_{zw}\,a^\dagger_z(0)\,a_{w}(0)
- \sum_{zw}L^*_z\Big(\frac{K}{\openone-S^*}\Big)_{zw}a_{w}(0) \\
{}+ \sum_{zw}\Big(\frac{K}{S-\openone}\Big)_{zw}L_z\,a^\dagger_{w}(0)
+ H + \sum_{zw}L^*_z\Big(\frac{K-\sin K}{4\big(\sin(K/2)\big)^2}\Big)_{zw}L_{w},
\end{multline}
which allows to read more explicitly the interaction between the systems.
In the special case $L=0$ we have
\begin{equation}\label{L=0}
H_T= E_0 + \sum_{zw}K_{zw}a^\dagger_z(0)\,a_{w}(0)+ H,
\end{equation}
while for $K=0$, i.e.\ $S=\openone$, we get
\begin{equation}\label{S=1}
H_T= E_0 - \rmi\sum_{z}L^*_z\,a_{z}(0) + \rmi\sum_{z}L_z\,a^\dagger_{z}(0) + H.
\end{equation}

As initial state let us take $|e(v)\rangle\langle e(v)|\otimes \rho_0$, where
$\rho_0\in\Sscr(\Hscr)$ is the initial system state and $e(v)$ is the coherent vector in $\Kscr=\Gamma[L^2(\Rbb;\ms)]$ with argument $v$ in $L^2(\Rbb;\ms)$. At the end it will be possible to take $v$ only locally square integrable.

Then, thanks to the properties of the HP-equation, the dynamics of the reduced system state
\begin{multline}\label{red_st}
\eta(t):=\Tr_\Kscr\left\{U(t)\left(| e(v)\rangle\langle e(v)|\otimes \rho_0\right)U(t)^*\right\}
\\ {}=\Tr_\Kscr\left\{V(t)\left(| e(v)\rangle\langle e(v)|\otimes \rho_0 \right)V(t)^*\right\}
\end{multline}
is given\cite{Parthas92,Bar06} by the master equation $\dot \eta(t)=\Lcal(t)[\eta(t)]$ with Liouville operator
\begin{subequations}\label{Lop}
\begin{equation}
\Lcal(t)[\tau]=-\rmi [H(t),\tau]+ \sum_z\left(\tilde L_z(t)\tau\tilde L_z(t)^* -\frac 1 2 \left\{ \tilde L_z(t)^* \tilde L_z(t),\tau\right\}\right),
\end{equation}
\begin{equation}
\tilde L_z(t):=L_z +\sum_w \left(S_{zw}-\delta_{zw}\right)v_w(t),
\end{equation}
\begin{equation}
H(t):=H+\frac \rmi 2 \sum_{zw}\left[ \overline{v_z(t)} \left(S_{wz}^*+\delta_{zw}\right) L_w + \overline{v_z(t)} S_{zw} v_w(t)-\text{h.c.}\right].
\end{equation}
\end{subequations}
Of course the reduced evolution depends on the global dynamics \eqref{VHPeq} and on the environment initial state. But this correspondence is not injective at all, so that it is not enough to know the Liouvillian $\Lcal$ to know the system/environment interaction.

\subsection{From the HP-equation to the SSE\label{sec:toSSE}}

The fields which have already interacted with $\Hscr$ can be manipulated in various ways and then monitored continuously in time. In this way we avoid to further perturb the dynamics of $\Hscr$, but, at the same time, as we indirectly acquire information on its state, the dynamics of $\Hscr$ turns out to be conditioned by the observed output. In the typical case of quantum optics the system is a photoemissive source and the output fields are mixed up by means of beam splitters and optical fibers and detected by photon counters (direct, homodyne, heterodyne detection)\cite{Bar06}. In general, we identify a measurement in continuous time by a family of commuting selfadjoint field operators which can be chosen as follows.

The manipulation of the fundamental fields is represented by a unitary, possibly time dependent, matrix $u_{iz}(t)$,
\[
\sum_{i\in Z} \overline{ u_{iz}(t)}\, u_{iw}(t) =\delta_{zw}, \qquad \sum_{z\in Z} u_{iz}(t) \overline{ u_{jz}(t)}=\delta_{ij},
\]
and produces the new field operators
\begin{gather*} B_i(t):= \sum_{z\in Z} \int_0^tu_{iz}(s)\, \rmd A_z(s),
\\
\hat\Lambda_{ij}(t):=\sum_{z,w\in Z}\int_0^t \overline{u_{iz}(s)}\,u_{jw}(s)\,\rmd \Lambda_{zw}(s), \qquad i,j\in Z.
\end{gather*}
Then, set $\dim\ms=d+d'$, we choose as observables the commuting selfadjoint operators (interaction picture)
\begin{equation}\label{observables}
B_i(s)+B_i^\dagger(s), \quad \hat\Lambda_{kk}(s), \quad i=1,\ldots ,d, \ k=d+1,\ldots, d+d', \quad s\geq 0.
\end{equation}

The global evolution \eqref{VHPeq}, the environment initial coherent state $|e(v)\rangle \langle e(v)|$ and the observed fields \eqref{observables} together determine both the distribution of the output processes and the a posteriori dynamics of the system $\Hscr$, that is the evolution of $\Hscr$ as a function of the observed outputs; both of them depending on the system initial state $\rho_0$. As we observe a maximal family of compatible fields, the a posteriori evolution preserves the purity of the system states and thus the problem of dynamics and observation can be reduced to a classical linear SSE\cite{BarP96,BarGreg09}:
\begin{multline}\label{schrodinger}
\rmd \varphi (t) = \K(t)\varphi(t_-) \rmd t + \sum_{j=1}^{d}R_j(t) \varphi(t_-)\rmd W_j(t)
\\ {}
+ \sum_{k=1}^{d'}
\left[\left(\frac{J_k(t)}{\sqrt{\lambda_k}}-\openone\right)\varphi(t_-)\rmd N_k(t) + \frac{\lambda_k}2\, \varphi(t_-) \rmd t\right],
\end{multline}
\begin{equation}\label{Kop}
\K(t):=
-\rmi H_0(t) -\frac 1 2\sum_{j\in Z} R_j(t)^*R_j(t).
\end{equation}
\begin{subequations}\label{newop}
\begin{equation}
H_0(t):=H+\frac \rmi 2 \sum_{z,w\in Z}\left( \overline{v_z(t)} S_{wz}^* L_w -L_w^*S_{wz}v_z\right),
\end{equation}
\begin{equation}\label{RjJk}
R_j(t):= \sum_{z\in Z} u_{jz}(t) \left(L_z+\sum_{w\in Z}S_{zw}v_w(t)\right),
\qquad J_k(t):=R_{d+k}(t);
\end{equation}
\end{subequations}
the initial condition is $\varphi(0)=\psi _0 \in \Hscr$, $\norm{\psi_0}=1$.
Equation \eqref{schrodinger} is a stochastic differential equation for a $\Hscr$-valued stochastic process $\varphi(t)$ in a filtered probability space, say $\big(\Omega, \Fscr, (\Fscr_t), \Qbb\big)$, where $W_j$, $N_k$ are independent Wiener and Poisson processes, each $N_k$ with rate $\lambda_k$. The solution $\varphi(t)$ is taken continuous from the right and $\varphi(t_-)$ in the right hand side means that the value of the solution is taken before of the possible jump at time $t$. The solution $\varphi(t)$ is a function of the initial condition $\psi_0$ and of the trajectories of the processes $W_j$ and $N_k$ up to time $t$.

Equation \eqref{schrodinger} can be translated in the language of stochastic processes $\sigma(t)$ taking values among positive operators on $\Hscr$. Indeed, if $A(t,s)$ is the fundamental solution of Eq.\ \eqref{schrodinger}, or the \emph{propagator} from time $s$ to $t$, taken $\rho_0\in\Scal(\Hscr)$, the stochastic process $\sigma(t):=A(t,0) \rho_0A(t,0)^*$  satisfies the linear stochastic master equation
\begin{multline}\label{lSME}
\rmd \sigma(t)=\Lcal(t)[\sigma(t_-)]\rmd t + \sum_{j=1}^{d}\bigl(R_j(t) \sigma(t_-)+\sigma(t_-) R_j(t)^* \bigr)\rmd W_j(t)\\ {}+
\sum_{k=1}^{d'}
\left[\left(\frac{J_k(t)\sigma(t_-)J_k(t)^*}{\lambda_k}-\sigma(t_-)\right)\bigl(\rmd N_k(t) - \lambda_k\,  \rmd t\bigr)\right],
\end{multline}
where $\Lcal(t)$ is the Liouville operator defined in Eqs.\ \eqref{Lop}.

Starting from Eq.\ \eqref{schrodinger} or Eq.\ \eqref{lSME} one can get both the distribution of the outputs and the a posteriori dynamics of $\Hscr$.

Of course, the joint distribution of the compatible field observables $B_i^\dagger(t)+ B_i(t)$ and $\hat\Lambda_k(t)$ is given by the Born rule based on their joint projection valued measure and the initial system/field state. Anyway, it can be obtained directly from Eq.\ \eqref{schrodinger} or Eq.\ \eqref{lSME}, as it is the joint distribution of the processes $W_j$, $N_k$ under the \emph{physical probability} on $(\Omega, \Fscr_T)$:
\begin{equation}\label{newprob}
\Pbb_T(\rmd
\omega)=p_T(\omega)\Qbb(\rmd \omega),\qquad p_t=\Tr\left\{\sigma(t)\right\} .
\end{equation}

Moreover, by defining $\rho(t):= \frac{\sigma(t)}{p_t}$ when $p_t>0$, and by taking an arbitrary state for $\rho(t)$ when $p_t=0$, we obtain the so called \emph{a posteriori state}, the conditional state to be attributed to the system, having observed the realization of \emph{all} the processes $W_j$ and $N_k$ up to time $t$. Correspondingly, let us call $\sigma(t)$ \emph{the non normalized a posteriori state}

In particular, regarding the distribution of the outputs, by Girsanov theorem we can say that under the physical probability $\Pbb_T$
\begin{equation}\label{newW}
\widehat W_j(t):=W_j(t)-\int_0^tm_j(s)\rmd s, \qquad m_j(t):= 2\RE \Tr\left\{ R_j(t)
\rho(t_-) \right\},
\end{equation}
$j=1,\ldots , d$, is a $d$-dimensional standard Wiener process, while $N_k(t)$ is a counting process of stochastic intensity $
\mu_k(t)=\Tr\left\{J_k(t)^*J_k(t) \rho(t_-)\right\}$.

As we observe the fields without introducing any new disturbance on $\Hscr$, we have that its a priori state, that is the mean of its a posteriori states, coincides with its reduced state \eqref{red_st} in absence of measurement:
\begin{equation}\label{mean=eta}
\eta(t)=\Ebb_{\Pbb_T}\left[\rho(t)\right]=\Ebb_{\Qbb}\left[\sigma(t)\right], \qquad t\in [0,T].
\end{equation}
Thus, the continuous measurement gives an unravelling (with a physical interpretation) to the open dynamics \eqref{Lop}. Of course, if we change the observed fields for a given global evolution and a given environmental initial state, we get a different unravelling of the same open evolution.

\subsection{Interacting and non interacting subsystems}

Let us finally consider a bipartite system $\Hscr=\Hscr_1\otimes\Hscr_2$ with its environment $\Kscr=\Gamma[L^2(\Rbb;\ms)]$ and their HP-evolution \eqref{VHPeq}. We are interested in the case of no direct interaction between the two subsystems $\Hscr_1$ and $\Hscr_2$, but, because of the common environment $\Kscr$, the two subsystems could have or not have an indirect interaction.

If the global Hamiltonian $H_T$ were bounded, we could say that $\Hscr_1$ and $\Hscr_2$ do not interact directly if the global Hamiltonian is
$$H_T=H_0+H_1+H_2+H_{01}+H_{02}$$
where $H_0=H_0^*\in\Lscr(\Kscr)$ is the free Hamiltonian of the environment, $H_1=H_1^*\in\Lscr(\Hscr_1)$ is the free Hamiltonian of $\Hscr_1$, $H_2=H_2^*\in\Lscr(\Hscr_2)$ is the free Hamiltonian of $\Hscr_2$, while $H_{01}\in\Lscr(\Kscr\otimes\Hscr_1)$ and $H_{02}\in\Lscr(\Kscr\otimes\Hscr_2)$ give the interaction, respectively, of $\Hscr_1$ with $\Kscr$ and of $\Hscr_2$ with $\Kscr$.

Analogously, dealing with HP-evolutions, we say that $\Hscr_1$ and $\Hscr_2$ do not interact directly if, in the heuristic representation \eqref{heuristichamiltonian} of the global Hamiltonian $H_T$, each one of the operators $K_{zw}$, $\sum_{w}\Big(\frac{K}{S-\openone}\Big)_{zw}L_w$ and $H + \sum_{zw}L^*_z\Big(\frac{K-\sin K}{4\big(\sin(K/2)\big)^2}\Big)_{zw}L_{w}$ is the sum of local operators. This property is independent of the basis $\{|z\rangle\}_{z\in Z}$ chosen in $\ms$.

In the case $L=0$, this means $H=H_1+H_2$, with $H_\ell=H_\ell^*\in\Lscr(\Hscr_\ell)$, and $K=K_1+K_2$, with $K_\ell=K_\ell^*\in\Lscr(\ms\otimes\Hscr_\ell)$.

In the case $K=0$, this means $H=H_1+H_2$, with $H_\ell=H_\ell^*\in\Lscr(\Hscr_\ell)$, and each $L_z=L_z^{(1)}+L_z^{(2)}$, with $L_z^{(\ell)}\in\Lscr(\Hscr_\ell)$.

An important subcase is when the subsystems $\Hscr_1$ and $\Hscr_2$ do not have any kind of interaction, either direct or indirect. In other words, this means that each subsystem $\Hscr_\ell$ has its own environment $\Kscr_\ell$ and that there is no interaction between $\Hscr_1$ and $\Kscr_1$ on one side and $\Hscr_2$ and $\Kscr_2$ on the other.
Thus, we say that $\Hscr_1$ and $\Hscr_2$ do not interact, either directly or indirectly, if there exists a decomposition $\ms=\ms_1\oplus\ms_2$, that is a decomposition $\Kscr=\Gamma[L^2(\Rbb;\ms)]=\Gamma[L^2(\Rbb;\ms_1)]\otimes\Gamma[L^2(\Rbb;\ms_2)]=\Kscr_1\otimes\Kscr_2$ such that, chosen a basis $\{|z\rangle\}_{z\in Z_1}$ in $\ms_1$ and a basis $\{|z\rangle\}_{z\in Z_2}$ in $\ms_2$ and considering the heuristic representation \eqref{heuristichamiltonian} of the global Hamiltonian in the basis $\{|z\rangle\}_{z\in Z_1\cup Z_2}$ in $\ms$, each addendum is an operator on $\Hscr_1\otimes\Kscr_1$ or on $\Hscr_2\otimes\Kscr_2$. This means that $K_{zw}$ belongs $\Lscr(\Hscr_\ell)$ when both $z,w\in Z_\ell$, while it is null otherwise, that $\sum_{w}\Big(\frac{K}{S-\openone}\Big)_{zw}L_w$ belongs to $\Lscr(\Hscr_\ell)$ when $z\in Z_\ell$, and that $H + \sum_{zw}L^*_z\Big(\frac{K-\sin K}{4\big(\sin(K/2)\big)^2}\Big)_{zw}L_{w}$ is the sum of local operators. This property is independent of the bases chosen in $\ms_1$ and $\ms_2$.

In the case $K=0$, this means $H=H_1+H_2$, with $H_\ell=H_\ell^*\in\Lscr(\Hscr_\ell)$, and $L_z\in \Lscr(\Hscr_1)$ for $z\in Z_1$, $L_z\in \Lscr(\Hscr_2)$ for $z\in Z_2$.

Let us remark that the Liouvillian \eqref{Lop} is not enough to understand whether the subsystems $\Hscr_1$ and $\Hscr_2$ do or do not interact.

\section{No direct or indirect interaction\label{sec:noint}}
Let us start by the last case presented in the previous section, when the two qubits do not interact either  directly or indirectly through a common bath, and let us study the role of a complete continuous measurement.
We consider only the case $S=\openone$, so that we need to take $Z=Z_1\cup Z_2$, $Z_1\cap Z_2=\emptyset$,
\begin{equation}\label{eq:LH}
L_z=\begin{cases} \hat L_z\otimes \openone & \text{ for } z\in Z_1, \\
\openone\otimes\hat L_z & \text{ for } z\in Z_2,
\end{cases} \qquad
H=H_1\otimes \openone+\openone \otimes H_2.
\end{equation}
As it will be useful in the following, from now on we give evidence to the tensor product structure of the various operators we need. Then, from Eqs.\ \eqref{Lop} we get the Liouville operator $\Lcal(t)= \Lcal_1(t)\otimes \openone+\openone \otimes \Lcal_2(t)$ with
\[
\Lcal_i(t)[\tau]:=-\rmi [H_i(t),\tau]+ \sum_{z\in Z_i}\left(\hat L_z\tau{\hat L_z}^* -\frac 1 2 \left\{ {\hat L_z}^* \hat L_z,\tau\right\}
\right),
\]
\[
H_i(t):=H_i+\rmi\sum_{z\in Z_i}\left( \overline{v_z(t)}\, \hat L_z- v_z(t){\hat L_z}^* \right).
\]
Recall that $v$ is the argument in the environment initial coherent state. Moreover, in the case of a complete observation, we obtain the SSE \eqref{schrodinger} with
$R_j(t)$ and $J_k(t)$ given by Eq.\ \eqref{RjJk}, $\K(t)=\K_1(t)\otimes \openone+\openone \otimes \K_2(t)
$,
\[
\K_i(t):=-\rmi H_i-\frac 1 2 \sum_{z\in Z_i}\left( \hat L_z^* \hat L_z +2 v_z(t) \hat L_z^* + \abs{v_z(t)}^2 \right).
\]

%New probability on $(\Omega, \Fscr_T)$: \ $\Pbb_T(\rmd
%\omega)=p_T(\omega)\Qbb(\rmd \omega)$, $p_t=\norm{\varphi(t)}^2 $.
%
%A posteriori state (complete observation): \quad $\displaystyle
%\psi(t)=\frac {\varphi(t)} {\norm{\varphi(t)}}
%$
%\[
%\rmd p_t =p_t\biggl\{\sum_jm_j(t)\rmd W_j(t)+\sum_k\left(\frac{\mu_k(t)}{\lambda_k}-1\right) \left(\rmd N_k(t)-\lambda_k\rmd t\right)\biggr\}
%\]
%\[
%m_j(t)= 2\RE \langle\psi(t) | R_j(t)
%\psi(t) \rangle
%\qquad
%\mu_k(t)=\norm{J_k(t) \psi(t)}^2
%\]
%\begin{multline*}
%p_t=\exp\biggl\{ \sum_j\biggl[
%\int_0^t m_j(s) \rmd W_j(s) -\frac 1 2 \int_0^t m_j(s)^2 \rmd s \biggr]\\ {}+ \sum_k\int_0^t\left( \lambda_k -\mu_k(s)\right) \rmd s
%\biggr\} \prod_{0<s\leq t} \prod_k \left(\frac {\mu_k(s)} {\lambda_k} \right)^{\Delta N_k(s)}
%\end{multline*}
%
%New Wiener process: \quad $\displaystyle
%\rmd \widehat W_j(t)=\rmd W_j(t)-m_j(t)\rmd t
%$
%
%\begin{equation}\label{rhot}
%\rho(t)=\Ebb_{\Pbb_T}\left[|\psi(t)\rangle \langle \psi(t)|\right]=
%\Ebb_{\Qbb}\left[|\varphi(t)\rangle \langle \varphi(t)|\right]
%\end{equation}

Let us start by considering a pure initial state $\rho_0=|\psi_0\rangle\langle \psi_0|$, so that $\sigma(t) = |\varphi(t)\rangle\langle \varphi(t)|$ and $p_t=\norm{\varphi(t)}^2$, cf.\ Eqs.\ \eqref{schrodinger}, \eqref{lSME}, \eqref{newprob}. Now the random a posteriori states are given by $\rho(t)=|\psi(t)\rangle\langle \psi(t)|$ with $\psi(t)=\varphi(t)/\norm{\varphi(t)}$ and the a priori states by $\eta(t)=\Ebb_{\Pbb_T}\left[\rho(t)\right]=\Ebb_{\Qbb}\left[\sigma(t)\right]$, see Sect.\ \ref{sec:QSDE}.

\subsection{The a posteriori concurrence}\label{genapost}

By the definition of concurrence in the case of pure states
\eqref{Cphi}, \eqref{Cpsi}, we can introduce the random \emph{a posteriori concurrence}
\begin{equation}\label{apostX}
C_{\rho(t)} \equiv C_{\psi(t)} = \frac{\abs{\chi_{\varphi(t)}}}{\norm{\varphi(t)}^2}
\end{equation}
and the \emph{mean a posteriori concurrence}
\begin{equation}\label{apostY}
\Ebb_{\Pbb_T}\left[C_{\psi(t)}\right]= \Ebb_{\Qbb}\left[\abs{\chi_{\varphi(t)}}\right],
\quad 0\leq t \leq T.
\end{equation}
By the definition of concurrence for mixed states \eqref{Cmixed} and of a priori states \eqref{mean=eta}, we get that the \emph{a priori concurrence} is bounded by the mean a posteriori concurrence:
\begin{equation} C_{\eta(t)}\leq \Ebb_{\Pbb_T}\left[C_{\psi(t)}\right].
\end{equation}

By the linear SSE and It\^o's formula we get the stochastic differential of $\chi_{\varphi (t)}$, which we shall need in the following,
\begin{equation}\label{dchi1}
\rmd \chi_{\varphi (t)} = \epsilon(t)\rmd t +\sum_{j=1}^d \ell_j(t) \chi_{\varphi(t)}\,\rmd W_j(t)
+\sum_{k=1}^{d'} \left[q_k(t)\rmd N_k(t)+\lambda_k \chi_{\varphi(t)}\,\rmd t\right],
\end{equation}
where
\begin{equation*}
\epsilon(t) :=
\Tr_{\Cbb^2}\left\{ \K_1(t)+ \K_2(t)\right\}\chi_{\varphi(t)}+ \sum_{j=1}^d  \langle \T R_j(t) \varphi(t) | \sigma _y \otimes \sigma _y
R_j(t) \varphi (t)  \rangle,
\end{equation*}
\begin{equation}
\ell_j(t) :=\sum_{z\in Z}u_{jz}(t)\left(\Tr_{\Cbb^2}\hat L_z+2v_z(t)\right),
\end{equation}
\[
q_k(t) := \frac 1 {\lambda_k}  \langle \T J_k(t) \varphi(t) | \sigma _y \otimes \sigma _y
J_k(t) \varphi (t)  \rangle -\chi_{\varphi (t)}.
\]

By writing
\begin{equation}\label{Lsigma}
\hat L_z=\sum_{i=1}^3 h_{zi} \sigma_i+ r_z,
\end{equation}
we get
\begin{equation}
\ell_j(t)=2\sum_z u_{jz}(t)\bigl(r_z+v_z(t)\bigr),
\end{equation}
\begin{multline}
\Tr_{\Cbb^2}\left\{ \K_1(t)+ \K_2(t)\right\}=-\rmi \Tr_{\Cbb^2}\left\{ H_1+ H_2\right\}
\\ {}- \sum_{z\in Z} \biggl\{ \sum_{i=1}^3 \abs{h_{zi}}^2+ \abs{r_z}^2 + \abs{v_z(t)}^2 + 2 v_z(t)\,\overline{r_z}\biggr\}.
\end{multline}
Let us stress that the operators $R_j(t)$ and $J_k(t)$ are not in general local operators, but sums of local operators. By this fact we cannot write in a more explicit form the coefficients $\epsilon(t)$ and $q_k(t)$.

\subsection{Only local detection operators} As already said, in this section we are considering only local operators in the dynamics: every qubit has its own environment and there is no direct nor indirect interaction between the two qubits. Now we consider the case in which also the detection operators are local, that is
\begin{equation}\label{localR}
R_j(t)= R^0_j(t)\otimes \openone \qquad \text{\textbf{or}} \qquad
R_j(t)= \openone \otimes R^0_j(t).
\end{equation}
This means that we observe separately the two environments.
With this further assumption, the stochastic differential \eqref{dchi1} becomes the closed equation
\begin{equation}\label{dchi2}
\rmd \chi_{\varphi(t)}=\chi_{\varphi(t)}\biggl( \kappa(t) \rmd t + \sum_j \ell_j(t)
\rmd W_j(t)+\sum_k\left(\frac {d_k(t)} {\lambda_k} -1 \right)\rmd N_k(t)\biggr),
\end{equation}
\[
\kappa(t) =
\Tr_{\Cbb^2}\left\{ \K_1(t)+ \K_2(t)\right\}+\sum_{k=1}^{d'}\lambda_k +  \sum_{j=1}^d
{\det}_{\Cbb^2} R_j^0(t),
\]
\begin{equation}\label{ljdk}
\ell_j(t) = \Tr_{\Cbb^2} R_j^0(t),
\qquad d_k(t)=
{\det}_{\Cbb^2} R_{d+k}^0(t).
\end{equation}

Equation \eqref{dchi2} can be explicitly solved and, by stochastic calculus, we get
\begin{equation}\label{meanC}
\Ebb_{\Pbb_T}\left[C_{\psi(t)}\right]= \Ebb_{\Qbb}\left[C_{\varphi(t)}\right]=C_{\psi_0}
\exp\biggl\{-\int_0^t c(s)\rmd s\biggr\},
\end{equation}
\[
c(t):=
\sum_{k=1}^{d'} \left(\lambda_k -  \abs{d_k(t)}\right)
-\frac 1 2 \sum_{j=1}^d
\left(\IM\ell_j(t)\right)^2- \RE\kappa(t).
\]
The first important result is that $c(t)$ does not depend on the initial state of the qubits, but only on the operators involved in the reduced dynamics and in the observation. This result is a slight generalization of the analogous one in Ref.\ \cite{VogS10}. By using \eqref{Lsigma},
we get, by straightforward calculations,
\begin{equation}\label{exprR}
R_j^0(t)=\sum_{i=1}^3 \tilde h_{ji}(t)\sigma_i+ \frac{\ell_j(t)}2, \qquad \tilde h_{ji}(t):=\sum_{z\in Z}u_{jz}(t)h_{zi},
\end{equation}
\begin{equation}\label{dkc}
d_k(t)= \frac{\ell_{d+k}(t)^2}4-\sum_{i=1}^3 \tilde h_{\left(d+k\right)i}(t)^2, \qquad c(t)=\sum_{j\in Z} c_j(t),
\end{equation}
\begin{equation}\label{c_j1}
c_j(t)=2\sum_{i=1}^3\left(\RE \tilde h_{ji}(t)\right)^2\geq0, \qquad j\leq d,
\end{equation}
\begin{equation}\label{c_j2}
c_j(t)=\frac 1 4\abs{\ell_j(t)}^2 - \abs{d_{j-d}(t)} + \sum_{i=1}^3\abs{\tilde h_{ji}(t)}^2 \geq 0,\quad j>d.
\end{equation}

By the fact that $c$ does not depend on the initial state of the qubits we can extend the result to the case of an initial mixed state $\rho_0$ and we get
\begin{equation}\label{C_eta,rho}
C_{\eta(t)}\leq \Ebb_{\Pbb_T}\left[C_{\rho(t)}\right]=C_{\rho_0}
\exp\biggl\{-\int_0^t c(s)\rmd s\biggr\};
\end{equation}
we assume always complete observation. Note that the mean a posteriori concurrence is non increasing. Moreover,
\begin{equation}\label{decay} \int_0^{+\infty}
c(s)\rmd s= +\infty \ \Rightarrow \ \lim_{t\to
+\infty}\Ebb_{\Pbb_t}\left[C_{\rho(t)}\right]=0,
\end{equation}
and, if $ c(t)=c>0$, the mean
a posteriori concurrence decreases  exponentially.

For what concerns the a priori states $\eta(t)$, when the master equation involves only local operators, one can have
the phenomenon of entanglement sudden death (ESD)\cite{Ors11,VogS10}. Note that no revival is possible for the concurrence of $\eta(t)$ due to the
bound given by the mean a posteriori concurrence \eqref{C_eta,rho}.

Also the a posteriori concurrence, without the mean, can be studied.
From the SDEs \eqref{schrodinger} for $\varphi(t)$ and \eqref{dchi2} for $\chi_{\varphi(t)}$, we can compute the stochastic differential of the concurrence $C_{\psi(t)}=\abs{\chi_{\varphi(t)}}\big/\norm{\varphi(t)}^2$; in terms of the new Wiener process \eqref{newW}, the final result is the closed SDE
\begin{multline}\label{SDEforC}
\rmd C_{\psi(t)}=C_{\psi(t)}\biggl\{ \sum_{j=1}^d \left[n_j(t) \rmd \widehat W_j(t) -c_j(t)\right]
\\ {}+\sum_{k=1}^{d'}\left[\left(\frac{\abs{d_k(t)}}{\mu_k(t)} -1 \right) \left(\rmd N_k(t)- \mu_k(t)
\,\rmd t\right)-c_{d+k}(t)\,\rmd t\right]\biggr\},
\end{multline}
where the $c_j(t)$ are given by Eq.\ \eqref{c_j1}, the $c_{d+k}(t)$ by Eq.\ \eqref{c_j2}, the $d_k(t)$ by Eq.\ \eqref{dkc} and
\begin{equation}\label{n_j}
n_j(t):=\RE\ell_j(t) - m_j(t)=-2\sum_{i=1}^3\left(\RE \tilde h_{ji}(t)\right)\langle \psi(t)|\mathtt{s}_i\psi(t)\rangle,
\end{equation}
\[
\mu_k(t)=\norm{\left(\sum_{i=1}^3 \tilde h_{(k+d)\,i}(t)\mathtt{s}_i + \frac{\ell_{k+d}(t)}2\right)\psi(t)}^2,
\]
\[
\tilde h_{ji}(t)\mathtt{s}_i=\begin{cases}
\tilde h_{ji}(t)\,\sigma_i\otimes \openone & \text{ if } R_j(t)=R_j^0(t)\otimes \openone,\\
\tilde h_{ji}(t)\,\openone \otimes \sigma_i & \text{ if } R_j(t)=\openone\otimes R_j^0(t).
\end{cases}
\]
The solution of the SDE \eqref{SDEforC} is given by the stochastic exponential
\begin{multline}\label{C_psi}
C_{\psi(t)}=C_{\psi_0} \exp\biggl\{ \sum_{j=1}^d\biggl[
\int_0^tn_j(s)\, \rmd \widehat W_j(s) - \int_0^t \left(c_j(s) +\frac {n_j(s)^2} 2\right)\rmd s \biggr]
\\{}-
\sum_{k=1}^{d'}\int_0^t\bigl(c_{d+k}(s)+\abs{d_k(s)} -\mu_k(s)\bigr) \rmd s
\biggr\} \;\prod_{0<s\leq t} \prod_{k=1}^{d'} \abs{\frac {d_k(s)} {\mu_k(s)} }^{\Delta N_k(s)}.
\end{multline}

\subsubsection{Diffusive case} Here we consider the purely diffusive case ($d'=0$). Now, in Eq.\ \eqref{meanC} the decay intensity of the mean a posteriori concurrence is $c(t)=\sum_{j=1}^d c_j(t)$ with $c_j(t)$ given by Eq.\ \eqref{c_j1}, while the random a posteriori concurrence reduces to
\begin{equation}\label{Cpsi(t)}
C_{\psi(t)}=C_{\psi_0} \exp\biggl\{ \sum_{j=1}^d\biggl[
\int_0^tn_j(s)\, \rmd \widehat W_j(s) - \int_0^t \left(c_j(s) +\frac {n_j(s)^2} 2\right)\rmd s \biggr]\biggr\},
\end{equation}
with $n_j(t)$ given by Eq.\ \eqref{n_j}.
%\[
%\Ebb_{\Pbb_T}\left[C_{\psi(t)}\right]=C_{\psi_0} \exp\biggl\{ -\sum_{j=1}^d\int_0^t c_j(s) \,\rmd s \biggr\}.
%\]
Let us stress that neither $c_j$ nor $n_j$
depend on the trace of the operators $R_j^0(t)$.

Note that, while the a priori states $\eta(t)$ can suddenly loose any entanglement (ESD), this is a.s.\ impossible for the a posteriori state (with complete
observation).

In the particular case of all the $R_j^0$'s selfadjoint there is  decay of the a posteriori concurrence, but, thanks to the freedom in the choice of the matrix $u$, by a change of phase we can pass from this case to the case of all the $R_j^0$'s anti-selfadjoint, for which there is no decay for every initial qubit state ($n_j=c_j=0$). Therefore, without
changing the master equation, i.e.\ without changing the dynamical behaviour of the
concurrence of the a priori state, one gets the complete entanglement protection by the choice of a phase in the detection operators.
The case of all the $R_j^0$ anti-selfadjoint gives $\norm{\varphi(t)}=$ constant and the SSE describes two independent random unitary evolutions.

\subsubsection{Jump case} Let us consider the purely jump case, i.e.\ $d=0$ and
$J_k(t)= J_k^0(t)\otimes \openone$ or
$J_k(t)= \openone \otimes J_k^0(t)$,
for which we have $\Ebb_{\Pbb_T}\left[C_{\psi(t)}\right]=C_{\psi_0} \rme^{ -\int_0^t c(s) \,\rmd s }$,
\begin{equation*}
C_{\psi(t)}=C_{\psi_0} \exp\biggl\{-
\sum_{k=1}^{d'}\int_0^t\bigl(c_{k}(s)+\abs{d_k(s)} -\mu_k(s)\bigr) \rmd s
\biggr\}
\prod_{0<s\leq t} \prod_{k=1}^{d'} \abs{\frac {d_k(s)} {\mu_k(s)}}^{\Delta N_k(s)},
\end{equation*}
\[
d_k(t)={\det}_{\Cbb^2} J_k^0(t)=\frac{\ell_{k}(t)^2}4-\sum_{i=1}^3 \tilde h_{ki}(t)^2,
\qquad
\mu_k(t)=\norm{J_k(t)\psi(t)}^2,
\]
\begin{equation*}
c(t)=\sum_{k=1}^{d'} c_k(t), \qquad c_k(t)=\frac 1 4\abs{\ell_k(t)}^2 - \abs{d_k(t)} + \sum_{i=1}^3\abs{\tilde h_{ki}(t)}^2 \geq 0.
\end{equation*}

One can check that $c_k(t)=0$ if and only if $\IM\left(\overline{\tilde h_{kj}(t)}\, \tilde h_{ki}(t) \right) =0$, $\RE\left(\overline{\ell_{k}(t)}\, \tilde h_{ki}(t) \right) =0$, $i,j=1,2,3$. Again, in some cases, one can protect the entanglement by tuning the detection operators without changing the mean dynamics, for instance by changing the unitary matrix $u(t)$. Let us give some examples.

A jump operator such as $J_k^0=h \sigma_i+\ell/2 $ contributes\cite{VogS10} with \[ c_{k}=\abs{h}^2+\frac{\abs{\ell}^2}4
- \sqrt{\left(\abs{h}^2+\frac{\abs{\ell}^2}4\right)^2- \frac 1 2\left(\RE \overline h\, \ell\right)^2}; \]
note that this contribution is zero when $\RE \overline h\, \ell=0$, while its maximum contribution is $c_k=
\abs{h}^2+\abs{\ell}^2/4
- \sqrt{\abs{h}^4+\abs{\ell}^4/16}$, reached when $\RE \overline h\, \ell=\pm\abs{ h\, \ell}$.

The jump operator
$J_k^0=\alpha \sigma_\pm+\beta $ contributes\cite{VogS10} with $c_k=\abs{\alpha}^2/2$.

Let us consider the term
\begin{equation}\label{gammadelta}
\gamma_-\left(\sigma_-\bullet \sigma_+ -\frac 1 2 \left\{\sigma_+ \sigma_- ,\bullet\right\}\right)+\gamma_+\left(\sigma_+\bullet \sigma_- -\frac 1 2 \left\{\sigma_- \sigma_+ ,\bullet\right\}\right)
\end{equation}
in the Liouville operator with $\gamma_+\geq 0$, $\delta> 0$, $\gamma_-=\delta+\gamma_+$. Three different choices of detection operators, but which give rise to the same dissipative term \eqref{gammadelta} in the master equation, are:
\begin{enumerate}
\item
$J_-=\sqrt{\gamma_-}\, \sigma_-$ and $J_+=\sqrt{\gamma_+}\, \sigma_+$, which contribute to $c$ with $
\gamma_++\delta/2$;
\item
$J_1=\sqrt{\gamma_+}\, \sigma_1$, $J_2=\sqrt{\gamma_+}\, \sigma_2$, $J_3=\sqrt{\delta}\, \sigma_-$, which contribute to $c$ with  $\delta/2$;
\item
$J_1= \frac 1{\sqrt{2}}\left(\sqrt{\gamma_+}\, \sigma_++ \sqrt{\gamma_-}\, \sigma_-\right)$, $J_2= \frac 1{\sqrt{2}}\left(\sqrt{\gamma_+}\, \sigma_+- \sqrt{\gamma_-}\, \sigma_-\right)$, which contribute to $c$ with $\frac 1 2 \left(\sqrt{\gamma_-}-\sqrt{\gamma_+}\right)^2$.
\end{enumerate}
Note that $\frac 1 2 \left(\sqrt{\gamma_-}-\sqrt{\gamma_+}\right)^2\leq \frac \delta 2 \leq \gamma_++ \frac \delta 2$. Given the dissipative term \eqref{gammadelta} in the Liouville operator, the choice (3) is the best one to slow down the disentanglement \cite[Eq.\ (19)]{VogS10}.

For what concerns the random a posteriori concurrence, if $C_{\psi(0)}>0$, $C_{\psi(t)}$ can vanish only if $d_k(s)\equiv {\det}_{\Cbb^2} J_k^0(s) =0$ for some $k$ and some $s$, as in the case of $\sigma_\pm$. In the jump case we can have ESD for some trajectories, eventually for all trajectories. The exponential decay of the mean concurrence is due to the randomness of the time of death.

\subsection{An example with general detection operators}\label{ex1}

Let us now consider a concrete model of non interacting qubits plus an environment. We want to show, in a very simple model, how the mere choice of the detection operators changes the behaviour of the a posteriori concurrence and how much this behaviour is different from the one of the a priori concurrence. The Liouville operator is fixed, but different choices of detection operators are studied.

As staring point \eqref{eq:LH} we take $Z=\{1,2\}$, $v(t)=0$,
\[
L_1=\hat L_1\otimes \openone, \quad L_2=\openone\otimes\hat L_2 , \quad \hat L_1=\hat L_2= \sqrt{\frac \gamma 2}\,\sigma_x, \quad \gamma>0,
\]
\[
H_1=H_2=\frac {\omega_0} 2 \, \sigma_z , \qquad \omega_0 \in \Rbb.
\]
The Liouville operator turns out to be
\[
\Lcal=\Lcal_0\otimes \openone+ \openone\otimes \Lcal_0, \qquad \Lcal_0[\tau]= -\rmi \, \frac {\omega_0} 2 \left[ \sigma_z,\,\tau\right]-\frac \gamma 4 \left[
\sigma_x,\left[\sigma_x,\,\tau \right]\right];
\]
we can also write
\begin{multline}\label{Lxx}
\Lcal[\eta]=-\rmi \,
\frac {\omega_0}
2 \left[ \sigma_z\otimes \openone + \openone\otimes\sigma_z,\, \eta\right] - \gamma \eta
\\ {}+\frac \gamma 2 \left(\sigma_x\otimes \openone\,\eta
\,\sigma_x\otimes \openone+ \openone\otimes\sigma_x \,\eta \,
\openone\otimes\sigma_x\right).
\end{multline}

The master equation with Liouville operator \eqref{Lxx} with $\omega_0\neq 0$ has a unique equilibrium state given by $\eta_{\mathrm{eq}}=\openone/4$. When $\omega_0= 0$, we have more equilibria, the statistical operators which are diagonal in the canonical basis. In any case the equilibrium states are separable.

\subsubsection{Concurrence of the a priori state}

If one writes down the master equation with Liouville operator \eqref{Lxx}, one sees
that it decomposes in subsystems of equations which can be solved analytically.
However, to simplify the analysis of the dynamics and the computation of the
concurrence, it is worthwhile to consider the subclass of the ``X'' states given in Section \ref{sec:conc}.
By checking the master equation with generator \eqref{Lxx} in the canonical
basis, one can see that the class of X states is preserved.

%\smallskip

\paragraph{Case $\omega_0\neq 0$.} By the fact that there is a unique equilibrium
state proportional to the identity, we get
\[
\lim_{t\to +\infty}\rho_{23}(t)=\lim_{t\to +\infty}\rho_{14}(t)=0,
\qquad \lim_{t\to +\infty}\rho_{jj}(t)=\frac 1 4.
\]
Then, if the initial X state has positive concurrence, it exists a finite time
$t_D>0$ for which $C_{\rho(t_D)}=0$ and we have entanglement sudden death.

%\smallskip

\paragraph{Case $\omega_0= 0$.} In this case there is not a unique equilibrium
state. As we shall see, the a priori concurrence is always limited by the exponential decay
\eqref{limit1} (local detection operators, diffusive case); one can also check that this limit is saturated when the initial state
is a Bell state. But we can have also ESD; for instance, take as initial state
$\rho_0=|\psi_0\rangle \langle \psi_0|$, $\psi_0=\frac 1 {\sqrt{2}}
\left(|10\rangle+\rmi |01\rangle\right)=\frac{1+\rmi}2\left(|\beta_1\rangle +
\beta_2\rangle\right)$, which is again an X state. By solving the master
equation and computing the concurrence by formulae \eqref{CX} we find ESD at the
time $t_D=-\frac 1 \gamma \,\ln \left(\sqrt{2} -1\right)$; moreover, for $t\in
[0,t_D]$ the a priori concurrence is given by
$C_{\eta(t)}= \frac 1 2 \left(1+ \rme^{-\gamma t}\right)^2 -1$.

\subsubsection{Local detection operators}
We start by considering local detection operators. The unitary matrix $u$ which fixes the observed fields in Sect.\ \ref{sec:toSSE} is taken to be $u_{jz}=\delta_{jz}\, \rme^{\rmi \phi_j}$, $\phi_j\in [0,2\pi]$. Then, the detection operators \eqref{RjJk}, \eqref{localR} reduce to
\[
R_1=\sqrt{\frac \gamma 2 }\, \rme^{\rmi \phi_1}\sigma_x\otimes \openone,
\qquad R_2=\sqrt{\frac \gamma 2 }\, \rme^{\rmi \phi_2}\openone\otimes\sigma_x.
\]

%\smallskip

\paragraph{Diffusive case.} Let us start by an observation of homodyne/heterodyne type: $d=2$, $d'=0$. Then, by Eqs. \eqref{exprR}--\eqref{C_eta,rho}, \eqref{n_j}, \eqref{Cpsi(t)} we get the a posteriori concurrence
\[
C_{\psi(t)}=C_{\psi_0} \rme^{-ct}\exp\biggl\{ \sum_{j=1}^2\biggl[
\int_0^tn_j(s)\, \rmd \widehat W_j(s) - \frac 1 2\int_0^t n_j(s)^2\rmd s \biggr]\biggr\}
\]
and the mean a posteriori concurrence
$\Ebb_{\Pbb_T}\left[C_{\rho(t)}\right]=C_{\rho_0} \rme^{-ct}$,
where \[
0\leq c=\gamma \left[\left(\cos\phi_1\right)^2+\left(\cos\phi_2\right)^2\right]\leq 2\gamma,
\]
\[
n_1(t)=\sqrt{2\gamma} \cos\phi_1 \,\langle \psi(t)|\sigma_x\otimes
\openone\,\psi(t)\rangle,\] 
\[n_2(t)=\sqrt{2\gamma} \cos\phi_2 \,\langle
\psi(t)| \openone \otimes\sigma_x \,\psi(t)\rangle.
\]
The important feature of this model is that it shows the dependence on the measuring
phases: the decay constant $c$ can take any value in the closed interval
$[0,2\gamma]$. Note that $c$ does not depend on $\omega_0$.
Finally, by the bound \eqref{C_eta,rho} for the a priori concurrence, we get
\begin{equation}\label{limit1}
C_{\eta(t)}\leq C_{\rho_0}\rme^{-2\gamma t}.
\end{equation}

\paragraph{Jump case.} Now let us consider a counting observation, with the same detection operators: $d'=2$, $d=0$, $J_k=R_k$. From \eqref{C_psi} we can check that the a posteriori concurrence turns out to be non random and constant: $C_{\psi(t)}=C_{\psi_0}$.
This is due to the fact that the jump operators are proportional to local unitaries. Thus, any initial entanglement can be perfectly protected just by a proper monitoring of the environment. Let us stress that the a priori concurrence always vanishes for long times and sometimes even in a finite time.

\subsubsection{Non local detection operators}
We give now an example of detection with non local operators for the same non interacting qubits. Now we measure in a non local way the environments of the qubits, but we do not change the interaction with the environments and, thus, their a priori dynamics. We consider only the diffusive case ($d=2$, $d'=0$) and we take the unitary matrix $u$ of Section \ref{sec:toSSE} to be
\[
u=\frac 1 {\sqrt{2}}\begin{pmatrix}
\rme^{\rmi(\theta+\phi)}& \rme^{\rmi(\theta-\phi)}\\ \rmi \rme^{\rmi(\theta+\phi)}&-\rmi \rme^{\rmi(\theta-\phi)}
\end{pmatrix};
\]
then, we get
\[
R_{1}=\frac{\rme^{\rmi\theta}\sqrt{\gamma }}{2}\left( \rme^{\rmi\phi}\sigma_x \otimes \openone
+\rme^{-\rmi\phi}\openone \otimes \sigma_x\right),\] \[
R_{2}=\frac{\rme^{\rmi\theta}\sqrt{\gamma }}{2}\left( \rmi\rme^{\rmi\phi}\sigma_x \otimes \openone
- \rmi\rme^{-\rmi\phi}\openone \otimes \sigma_x\right).
\]

By particularizing the general formulae of Sect.\ \ref{genapost} we obtain that the stochastic differential of $\chi_{\varphi(t)}$ does not contain the white noise term and we have
\begin{equation}\label{CD1}
\dot
\chi_{\varphi(t)}= -\gamma \chi_{\varphi(t)}+\gamma \rme^{2\rmi \theta} \Dcal(t), \qquad \Dcal(t) := \langle \T \varphi(t) | \sigma _z \otimes \sigma _z \varphi(t) \rangle.
\end{equation}
Again by stochastic differentiation, we get
\begin{gather}\label{CD2}
\dot \Dcal(t)= \gamma \rme^{2\rmi \theta} \chi_{\varphi(t)}-\gamma \Dcal(t)- \rmi \omega_0
\Ecal(t),
\\ \notag
\Ecal(t) := \langle \T \varphi(t) | \left(\sigma _z \otimes \openone +\openone
\otimes \sigma _z \right)\varphi(t) \rangle.
\end{gather}
By differentiation of $\Ecal$ we get more
complicated expressions, including terms with stochastic differentials. Anyway, from Eqs.\
\eqref{CD1}, \eqref{CD2} we obtain
\begin{equation}\label{CpmD}
\chi_{\varphi(t)}\pm \Dcal(t)= \rme^{-\gamma_\pm t} \bigl( \chi_{\psi_0}\pm \Dcal(0)\bigr)
\mp \rmi \omega_0 \int_0^t \rme^{-\gamma_\pm \left(t-s\right)} \Ecal(s)\rmd s,
\end{equation}
\[
\gamma_\pm:=\gamma \left(1\pm \rme^{2\rmi \theta}\right).
\]

In this model one can have a variety of behaviours
for the mean concurrence, such as revivals and creation of concurrence in the long
run. Let us see this in the simplest case.

\paragraph{The case $\omega_0=0$.} In this case we have
\begin{equation}
\chi_{\varphi(t)}= \frac 1 2 \, \rme^{-\gamma_+ t} \bigl( \chi_{\psi_0}+\Dcal(0)\bigr)
+ \frac 1 2 \, \rme^{-\gamma_- t} \bigl( \chi_{\psi_0}-\Dcal(0)\bigr).
\end{equation}
Being non random, by Eqs.\ \eqref{apostX}, \eqref{apostY}, we get
$\Ebb_{\Pbb_T}\left[C_{\psi(t)}\right]= \abs{\chi_{\varphi(t)}}$,
for all $T\geq t $.

If $\rme^{2\rmi \theta}\neq \pm 1$, we get $\RE \gamma_\pm>0$. Then, the mean a
posteriori concurrence decays exponentially at long times, but, depending on the initial state of the qubits, it can have also
revivals. For instance, by taking $\psi_0$ such that $\chi_{\psi_0}=0$ and $\Dcal(0)\neq 0$, we have
\[
\abs{\chi_{\varphi(t)}}=\frac 1 2 \abs{\Dcal(0)}\abs{\rme^{-\gamma_+ t} -\rme^{-\gamma_- t}}.
\]

If $\rme^{2\rmi \theta}= 1$, we get $\gamma_+=2\gamma$, $\gamma_-=0$ and
\begin{equation*}
\abs{\chi_{\varphi(t)}}= \frac 1 2 \abs{ \rme^{-2\gamma t} \bigl( \chi_{\psi_0}+\Dcal(0)\bigr)
+  \bigl( \chi_{\psi_0}-\Dcal(0)\bigr)}.
\end{equation*}
So, depending on the initial state of the qubits, some concurrence can survive (entanglement protection) or can be created in the long run (entanglement generation). The case $\rme^{2\rmi \theta}= -1$ is similar.

\section{An example with indirect interaction \label{sec:jump}}

In this last section we consider the case of indirect interaction between two qubits and, by means of an explicit model, we show that a very extreme scenario can occur: the interaction with the environment completely destroys any entanglement between the qubits, if no measurement is performed, while the same interaction generates maximally entangled states, independently of the initial state of the qubits, if the environment is simply continuously monitored after the interaction. Indeed, while in the long run the a priori state of the qubits becomes maximally chaotic, and thus separable, their a posteriori state becomes maximally entangled for every output of the continuous measurement.

We consider a couple of qubits $\Hscr=\Hscr_1\otimes\Hscr_2$ interacting with a sort of continuous flow $\Kscr=\Gamma[L^2(\Rbb;\ms)]$ of quadruples of qubits $\ms=\ms_1\otimes\ms_2\otimes\ms_1'\otimes\ms_2'$.

Let us denote by $\{|i\rangle\}_{i=0,1}$ the canonical basis in $\Hscr_1=\Hscr_2=\ms_1=\ms_2=\ms_1'=\ms_2'=\Cbb^2$ and then let us introduce the flip operator $F_\ell$ in $\ms_\ell\otimes\Hscr_\ell$:
$$F_\ell=\sum_{ij}|ij\rangle\langle ji|=F_\ell^*=F_\ell^{-1}=\rme^{-\rmi\frac{\pi}{2}(F_\ell-1)}.$$

Let us choose in $\ms$ the basis generated by the Bell bases in $\ms_1\otimes\ms_2$ and in $\ms_1'\otimes\ms_2'$, that is $\{|\beta_x\otimes\beta_{x'}\rangle\}_{x,x'=0,\ldots,3}$.

We consider the HP evolution \eqref{VHPeq} generated by the interaction $H=0$,
\begin{equation*}
L=0,\qquad
S=F_1\,F_2=F_2\,F_1=\rme^{-\rmi\frac{\pi}{2}(F_1+F_2-2)}, \quad K=\frac{\pi}{2}(F_1+F_2-2),
\end{equation*}
where every operator is identified with its natural extension. Roughly speaking, when a quadruple of qubits $\ms$ belonging to the continuous flow interacts with the couple of interest $\Hscr$, the first two qubits of the quadruple $\ms_1\otimes\ms_2$ exchange their joint state with $\Hscr$, while the other two qubits $\ms_1'\otimes\ms_2'$ are simple witnesses. Then
$$S_{(xx')(yy')}=\Tr_{\ms}\left[\big(|\beta_y\otimes\beta_{y'}\rangle\langle\beta_x
\otimes\beta_{x'}|\otimes\openone_\Hscr\big)S\right]=|\beta_y\rangle\langle\beta_x|
\,\delta_{x'y'}$$
and the Hudson-Parthasaraty equation is
\begin{equation*}
\rmd V(t)= \sum_{xyx'}\Big(|\beta_y\rangle\langle\beta_x|-\delta_{xy}\Big)V(t)\,\rmd\Lambda_{(xx')(yx')}(t).
\end{equation*}
%with solution
%\begin{multline*}
%V(t)=\overleftarrow{\exp}\Bigg\{-\rmi\int_0^t\sum_{xx'yy'}K_{(xx')(yy')}a^\dagger_{xx'}(s)a_{yy'}(s)\,\rmd s\Bigg\}\\
%=\overleftarrow{\exp}\Bigg\{-\rmi\frac{\pi}{2}\int_0^t\bigg[\sum_{xyx'}
%\Tr_{\ms_1}\{|\beta_y\rangle\langle\beta_x|\}\otimes\openone_{\Hscr_2}\,a^\dagger_{xx'}(s)\,a_{yx'}(s)\\
%+\sum_{xyx'}\openone_{\Hscr_1}\otimes\Tr_{\ms_2}\{|\beta_y\rangle\langle\beta_x|\}\,a^\dagger_{xx'}(s)
%\,a_{yx'}(s)\\
%-2\sum_{xx'}a^\dagger_{xx'}(s)\,a_{xx'}(s)\bigg]\,\rmd s\Bigg\}
%\end{multline*}
Therefore, there is no direct interaction between $\Hscr_1$ and $\Hscr_2$ as $K=\frac{\pi}{2}(F_1+F_2-2)$ with $F_1$ involving only $\Hscr_1$ and $F_2$ involving only $\Hscr_2$.
Let us remark that this is just one of those cases where the whole interaction is encoded in the domain of the global Hamiltonian $H_T$. Indeed,\cite{Greg01} $H_T$ is just an extension of the free field Hamiltonian $E_0$, re-restricted to the domain of the ``regular vectors'' $\Phi\in\Kscr\otimes\Hscr$ such that $a_{xx'}(0^-)\,\Phi=\sum_{yy'} S_{(xx')(yy')}\,a_{yy'}(0^+)\,\Phi$ for all $x,x'$.

For the environment we choose the initial pure coherent state $|e(v)\rangle\langle e(v)|$ with argument
$$v(t)=\sqrt{\frac{\nu}{4}}\sum_{x=0}^3|\beta_x\rangle\otimes|\beta_x\rangle\in\ms=
\Big(\ms_1\otimes\ms_2\Big)\otimes\Big(\ms_1'\otimes\ms_2'\Big), \qquad \forall0\leq t\leq T,$$
where $\nu$ is a positive parameter and $T>0$ is our arbitrary time horizon. Roughly speaking, even if the qubits $\ms_1'$ and $\ms_2'$ are not involved in the interaction with $\Hscr_1$ and $\Hscr_2$, they are initially entangled with the qubits $\ms_1$ and $\ms_2$ which exchange their state with $\Hscr_1$ and $\Hscr_2$.

Then, if $\rho_0$ is the system initial state, its reduced state at time $t$ is
\[\eta(t)=\Tr_\Kscr\Big[U_t\,|e(v)\rangle\langle e(v)|\otimes\rho_0\,U^*_t\Big]=\rme^{\Lscr t}\rho_0, \]%\quad %\forall0\leq t\leq T,$$
where
$$\Lcal\eta=\nu\,\frac{\Tr\eta}{4}\,\openone-\nu\,\eta,$$
so that
$$\eta(t)=\rho_0\rme^{-\nu t}+\frac{\openone}{4}(1-\rme^{-\nu t})\to\frac{\openone}{4},\qquad \text{for }t\to\infty,$$
and the state of $\Hscr$ becomes maximally chaotic and any entanglement between $\Hscr_1$ and $\Hscr_2$ is destroyed by the interaction with the common bath.

The a priori concurrence goes to 0 at least exponentially,
\[
C_{\eta(t)}\leq C_{\rho_0}\,\rme^{-\nu t}\to0,\qquad \text{for }t\to\infty,
\]
and, depending on the system initial state $\rho_0$, we can even assist to entanglement sudden death.

This can be verified by considering an X state as initial state. Indeed, if $\eta_{ij}$ are the matrix elements of $\rho_0$ with respect to the computational basis \eqref{cbas}, we find
\begin{subequations}
\begin{gather*}
C_{\eta(t)}=2\max \left\{ 0,\, C_1(t),\, C_2(t) \right\},\\
C_1(t)=\abs{\eta_{23}}\rme^{-\nu t} - \sqrt{\Big(\eta_{11}\rme^{-\nu t}+\frac{1}{4}(1-\rme^{-\nu t})\Big)\Big(\eta_{44}\rme^{-\nu t}+\frac{1}{4}(1-\rme^{-\nu t})\Big)},\\
C_2(t)=\abs{\eta_{14}}\rme^{-\nu t} - \sqrt{\Big(\eta_{22}\rme^{-\nu t}+\frac{1}{4}(1-\rme^{-\nu t})\Big)\Big(\eta_{33}\rme^{-\nu t}+\frac{1}{4}(1-\rme^{-\nu t})\Big)}.
\end{gather*}
\end{subequations}
By the fact that
\[
\lim_{t\to +\infty}C_1(t)=\lim_{t\to +\infty}C_2(t)=-\frac 1 4,
\]
if the initial X state $\rho_0$ has positive concurrence, it exists a finite time
$t_D>0$ for which $C_{\eta(t_D)}=0$. The death time $t_D$ can be explicitly computed. For example, if $\rho_0=|\beta_1\rangle\langle\beta_1|$, then $t_D=\frac{\ln3}{\nu}$.

Let us introduce now the continuous measurement. As a preliminary step, let us suppose we observe all the sixteen compatible processes of observables $\Lambda_{(xx')(xx')}(t)$. Roughly speaking, we count the quadruples of kinds $(xx')$ which have been through an interaction with the couple $\Hscr$ between time 0 and time $t$.
Then the corresponding linear stochastic master equation for the non normalized a posteriori state $\tilde\sigma(t)$ is
\begin{multline*}
\rmd\tilde\sigma(t)=\Lcal[\tilde\sigma(t_-)]\rmd t
\\ {}+\sum_{x,x'=0}^3\left(\frac{4\nu}{\lambda}\,|\beta_{x'}\rangle\langle\beta_x|\tilde\sigma(t_-)
|\beta_x\rangle\langle\beta_{x'}| - \tilde\sigma(t_-)\right)\left(\rmd N_{xx'}(t)-\frac{\lambda}{16}\rmd t\right)
\end{multline*}
in a probability space $(\Omega,\Fscr,\Fscr_t,N_{xx'}(t),\Qbb)$ where $N_{xx'}(t)$, $x,x'=0,\ldots,3$, are sixteen independent Poisson processes of rates $\lambda/16$ under $\Qbb$.

The definitive step is to consider the measurement of the (non maximal) family of the four compatible processes of observables
\begin{equation*}
\Lambda_{x'}(t)=
%\Lambda\Big(\pi_{(0,t)}\otimes\openone_{\ms_1}\otimes\openone_{\ms_2}\otimes|\beta_{x'}\rangle\langle\beta_{x'}|\Big)\\
%=\sum_{x=1}^4\Lambda\Big(\pi_{(0,t)}\otimes|\beta_{x}\rangle\langle\beta_{x}|\otimes|\beta_{x'}\rangle\langle\beta_{x'}|\Big)=
\sum_{x=0}^3\Lambda_{(xx')(xx')}(t).
\end{equation*}
Roughly speaking, we count the quadruples of qubits, with the second couple of kind $x'$, which have been through an interaction with the couple $\Hscr$ between time 0 and time $t$.
Then, by conditioning, we get the linear stochastic master equation for the non normalized a posteriori state $\sigma(t)$,
\begin{multline*}
\rmd\sigma(t)=\Lcal[\sigma(t_-)]\rmd t
\\ {}+ \sum_{x'=0}^3\left(\frac{\nu}{\lambda}\,\Big(\Tr\sigma(t_-)\Big)\,|\beta_{x'}\rangle\langle\beta_{x'}| - \sigma(t_-)\right)\left(\rmd N_{x'}(t)-\frac{\lambda}{4}\rmd t\right),
\end{multline*}
in a probability space $(\Omega,\Fscr,\Fscr_t,N_{x'}(t),\Qbb)$ where $N_{x'}(t)$, $x'=0,\ldots,3$, are four independent Poisson processes of rates $\lambda/4$ under $\Qbb$.

If $N(t)=\sum_{x'=0}^3 N_{x'}(t)$ denotes the total counts up to time $t$, $T_n$ denotes the arrival time of the count $n$ and if $X'_n$ denotes the mark of count $n$, the solution is
$$
\sigma(t)=\begin{cases}\rho_0\,\rme^{-\nu t+\lambda t},&\text{if }0\leq t<T_1,\\
|\beta_{X'_{N(t)}}\rangle\langle\beta_{X'_{N(t)}}|\,\rme^{-\nu t+\lambda t}\,\left(\frac{\nu}{\lambda}\right)^{N(t)},&\text{if }t\geq T_1.\end{cases}
$$
Then, under the physical probability $\Pbb_T(\rmd \omega)=\Tr\left\{\sigma(T)\right\}\Qbb(\rmd \omega)$, the four counting processes $N_{x'}(t)$ are four independent Poisson processes of rates $\nu/4$, which depend on the environment initial state, and the a posteriori state is
$$
\rho(t)=\begin{cases}\rho_0,&\text{if }0\leq t<T_1,\\
|\beta_{X'_{N(t)}}\rangle\langle\beta_{X'_{N(t)}}|,&\text{if }t\geq T_1.\end{cases}
$$

Roughly summarizing, a flow of quadruples of qubits $\ms$ interacts with the two qubits $\Hscr$. Actually only the couple $\ms_1\otimes\ms_2$ interacts by exchanging its state with $\Hscr$, while $\ms_1'\otimes\ms_2'$ is a simple witness which is, nevertheless, initially entangled with $\ms_1\otimes\ms_2$. As a result, the couple $\Hscr$ becomes entangled with the last couple $\ms_1'\otimes\ms_2'$ with which has interacted. By counting the quadruples gone through an interaction with $\Hscr$ and measuring the projection valued measure $\{|\beta_{x'}\rangle\langle\beta_{x'}|\}_{x'=0}^3$ on $\ms_1'\otimes\ms_2'$, we get an output with the distribution of a marked Poisson process and, at every count, the a posteriori state of $\Hscr$ jumps into the Bell state labelled by the corresponding mark $X'$.

We can also compute the random a posteriori concurrence
$$
C_{\rho(t)}=\begin{cases}C_{\rho_0},&\text{if }0\leq t<T_1,\\
1,&\text{if }t\geq T_1,\end{cases}
$$
and we find that the a posteriori concurrence goes to 1, both almost surely and in the mean,
\begin{gather*}
C_{\rho(t)}\to1,\qquad \text{for }t\to\infty, \qquad \text{$\Pbb$-a.s.}, \qquad \forall \rho_0,\\
\Ebb_\Pbb[C_{\rho(t)}]=1-(1-C_{\eta(0)})\rme^{-\nu t}\to1,\qquad \text{for }t\to\infty, \qquad \forall \rho_0,
\end{gather*}
while the a priori concurrence goes to 0,
$$
\Ebb_\Pbb[C_{\rho(t)}]\geq C_{\eta(t)}\to0,\qquad \text{for }t\to\infty.
$$

Therefore, while any entanglement between the qubits is a priori destroyed by the interaction with the common bath, at the same time it is enough to monitor the bath in a proper way to get a maximal creation of the a posteriori entanglement, for any initial state of the qubits.

\end{document}